\documentclass[12pt]{article}

\usepackage{amsmath}
\usepackage{amssymb}
\usepackage{graphicx}
\usepackage{color}
\usepackage{amssymb,amsfonts,amsmath}
\begin{document}

\begin{flushleft}
{\Large
\textbf{Tracing the Progression of Retinitis Pigmentosa via Photoreceptor Interactions}
}
~\\
Erika T.~Camacho$^{1}$,\\
Division of Mathematical \& Natural Sciences, Arizona State University, 4701 W. Thunderbird Rd, Glendale, AZ, 85306, USA, and
and Mathematical, Computational and Modeling Sciences Center, Arizona State University, P.O. Box 871904, Tempe, AZ, USA 85287-1904, erika.camacho@asu.edu;\\
Stephen Wirkus\\
Division of Mathematical \& Natural Sciences, Arizona State University, 4701 W. Thunderbird Rd, Glendale, AZ, 85306, USA, and 
and Mathematical, Computational and Modeling Sciences Center, Arizona State University, P.O. Box 871904, Tempe, AZ, USA 85287-1904,
swirkus@asu.edu;\\ 
\textbf{1} Corresponding author, tel: 602-543-8156, fax: 602-543-6073,
\\
\end{flushleft}

\begin{abstract} 
Retinitis pigmentosa (RP) is a group of inherited degenerative eye diseases characterized by mutations in the genetic structure of the photoreceptors that  leads to the premature death of both rod and cone photoreceptors.
Defects in particular genes encoding proteins that are involved in either the photoreceptor structure, phototransduction cascades, or visual cycle
are expressed in the rods but ultimately affect both types of cells. 
RP is ``typically'' manifested by a steady death of rods followed by a period of stability in which cones survive initially and then inevitably die too.  In some RP cases, rods and cones die off simultaneously or even cone death precedes rod death (reverse RP).
The mechanisms and factors involved in the development of the different types of RP are not well understood nor have researchers been able to provide more than a limited number of short-term therapies.
In this work we trace the progression of RP to complete blindness through each subtype via bifurcation theory.  We show that the evolution of RP from one stage to another often requires the failure of multiple components.
Our results indicate that a delicate balance between the availability of nutrients and the rates of shedding and renewal of photoreceptors is needed at every stage of RP to halt its progression. This work provides a framework for future physiological investigations potentially leading to long-term targeted multi-facet interventions and therapies dependent on the particular stage and subtype of RP under consideration. The results of this mathematical model may also give insight into the progression of many other degenerative eye diseases involving genetic mutations or secondary photoreceptor death and potential ways to circumvent these diseases.
\end{abstract}

{\bf keywords: retinal degeneration, rod-cone RP, reverse RP, bifurcation set, RdCVF}

\section*{Introduction}
\subsubsection*{Background of Retinitis Pigmentosa}

Retinitis Pigmentosa (RP) is a heterogeneous group of inherited disorders that affects approximately 1 in 4000 individuals (nearly 1.5 million) worldwide \cite{Shastry,Shintani,MohanSaid}.  This group of disorders is often classified as {\it primary} or {\it syndromic} RP.  In the case of syndromic RP, which account for about 20-30\% of all cases, the patients have an associated non-ocular disease with RP with abnormalities in one or more organs \cite{Hartong,Shintani}.  In contrast, primary RP involves only the eye.  RP is primarily genetically programmed with more than 45 genes associated with this disease identified to date.  These genes account for only about  60\% of all cases.  RP may be inherited from one or both parents (autosomal-recessive RP, autosomal-dominant, or X-linked RP, which account for approximately 20-30\%, 15-20\%, and 6-10\% of the cases, respectively) or may be caused by a new mutation even though there may be no family history of the disease \cite{Hartong,Shintani,Neidhardt,Bowne,Shastry,Busskamp}. Less common modes of inherited RP include digenic and mitochondrial \cite{Shintani}.

RP has been associated with mutations or errant genes encoding for proteins that are essential for rod survival and certain genetic products that are damaging to the rods \cite{Shintani,Hartong,Busskamp,Shen,Chalmel}.   Even though mutations are normally only expressed in the rods their effects ultimately cascade to the cones.  People usually are diagnosed only when they experience serious vision problems making it difficult to decipher the mechanism of development and the channels by which RP develops.  Adding to this challenge is the variability of its progression and its time of onset within each individual \cite{Shen,Wong,Hamel}.  The sequences of events (i.e., clinical manifestations) of the various sub-types of RP are well-documented \cite{Shintani,Phelan,Hamel} but this sequential progression varies from individual to individual.

RP is often classified into various sub-types depending on the location of the vision field that is affected.  The most common clinical manifestation is rod-cone RP and is characterized by difficulty seeing at night, followed by night blindness (nyctalopia), narrowing of the mid-peripheral field of vision (where the rods are more prevalent), and ultimately a loss of central vision\cite{Chalmel,Hartong,MohanSaid}.  Night blindness is the hallmark of 
rod-cone RP but given that rod function is primarily affected, this is of no surprise since rods contribute solely to night vision. Even though rod-cone RP has a better overall prognosis and slower progression in comparison with other sub-types of RP, the range of the onset of clinical rod-cone RP to its noticeable changes observed by the patient can vary widely in all cases.
In this most common form of RP, the causal mutations affect genes exclusively expressed in the rod cells in an overwhelming number of cases and rod death far exceeds and precedes cone death \cite{Shintani}.  Less common forms of RP see an equal loss of rods and cones or even a cone death exceeding rod death (cone-rod RP or reverse RP) \cite{Hartong,Shintani}.  Cone-rod dystrophy (reverse RP) is associated with a cone loss preceding rod loss and as a consequence reduction of the central vision, acuity, and color vision are manifested early in the progression of the disease, followed by reduction of the mid-peripheral  visual field and night blindness at later stages.  However, unlike the rod-cone RP where patients' color vision remains good until the central vision is affected at levels of 20/40 or worse, patients with cone-rod RP experience deterioration of their color vision, ability to adapt to light and day vision very early in the disease as well as the loss of normal cone functioning, which leads to photophobia.

\subsubsection*{Interactions of Photoreceptors}
Research that might shed light on improving the photoreceptors longevity has been paramount to physiological, anatomical, and medical research in the last 2 decades due to the prominent role that these cells play in the vision process \cite{Stone}.  The photoreceptors are the rods and cones, located in the retina, that convert light energy into nerve impulses through the process of visual transduction\cite{Keener}.  The rods and cones are specialized for different aspects of vision: the rods and cones are different in shape (cylindrical vs.~conical), contain different types of photopigment, and are distributed differently throughout the retina\cite{text}.  The reduction of normal light vision, visual acuity, detail perception, night vision, and narrowing of the peripheral fields can be caused by a depletion of cones and rods, respectively \cite{MohanSaid}.  The photoreceptors undergo tremendous oxidative stress over the course of a day and undergo a continuous cellular renewal and periodic shedding process of their outer segments (OS) in order to minimize depletion and help ensure proper functioning of these cells \cite{text,web,Kevany,Jonnal}.  Photoreceptors shed approximately the same amount of material that is renewed over the course of a day so that the length roughly remains constant, with complete renewal in rods and cones occurring from 11 days up to 3 weeks depending on the photoreceptor and its location within the retina \cite{text,Fisher,Tassi,LaVail1973,Young,Young1967,Anderson1980,Anderson,Guerin}.
The shedding and renewal of photoreceptors, their interactions, and circadian rhythms were previously examined and discussed in greater detail \cite{camacho, camacho2}. 

Much research has examined the processes of renewal and shedding within the photoreceptors and suggests the photoreceptors together with the adjacent retinal pigment epithelium (RPE) operate as a functional unit \cite{Tassi,Bhatt,ODay,Fisher,Hogan,Bok,Papermaster,Jonnal,YoungBok,
Young1967,Young,Besharse1988,Mukherjee,LaVail1973,Li,Strauss}.  The rod and cone cells undergo a daily rhythmic shedding in which their photoreceptor outer segments (POS) are phagocytized by the RPE. The RPE recycles parts of the shed POS for future use by the photoreceptors and contributes additional trophic and growth factors utilized in the renewal process of these cells \cite{Bhatt,MohanSaid,MohandSaid1998,Mukherjee}.  The RPE transports, secretes, and redelivers essential growth factors such as PEDF, CNTF, bFGF, Gas6, PDGF, and ATP, metabolites (including glucose, retinal, and omega-3 fatty acids), ions, and water to and from the photoreceptors \cite{Strauss,Murakami,Longbottom,LaVail1998,Li,Wenzel,Frasson}.  It is unclear if other essential proteins are linked to the interactions of these two retinal tissues but the interdependency of the RPE and photoreceptors is key to the proper functionality of the eye.  

Pivotal to the vision is also the connection and interplay between the rods and cones.  These cells have long been suspected of communicating directly or indirectly with each other \cite{MohandSaid1998}.  It has been recently shown that the rods produce a protein referred to as the Rod-derived Cone Viability Factor (RdCVF) that is essential for the functionality and survival of the cones
\cite{Leveillard,MohanSaid,Yang,Leveillard2010,Sahel,Hanein}.  
While the neural degeneration of the photoreceptors that typically results from genetic defects and mutations (such as RP) may disrupt and break the balanced shedding and renewal processes leading to shorter OS and, in some cases, the complete disappearance of the rods and cones \cite{MohanSaid,Pallikaris}, the amount of RdCVF appeared to be the same in both normal and mutated rods prior to their degeneration.  Additionally, a subsequent decrease in RdCVF was reported to be strongly correlated with the decrease in the number of rods during their degenerative process \cite{Yang}.  The effect of RdCVF appears to be independent of the casual mutation leading to rod degeneration \cite{Leveillard,Yang,Sahel}. Experiments have revealed that RdCVF had a significant cone rescuing effect, that its protective effects did not extend to degenerating rods, and that the presence of the rod cells is not necessary for cone survival as long as RdCVF is present \cite{Sahel,Leveillard,Yang,Leveillard2010,Hanein,camacho2}. Moreover, RdCVF was also shown to significantly preserve cone function, a very promising discovery since maintaining functional cones even when 95\% of the cones are gone may prevent blindness \cite{Sahel}.  
The discovery that rods produce RdCVF, which is essential for cone functionality and viability, and that their death deprives the cones from this element and triggers their degeneration has far-reaching implications.
It is yet unclear whether the RPE plays any role in the transport of this protein but it is clear that RdCVF provides an essential direct help from the rods to the cones without negatively affecting the rods.

\subsubsection*{Hypotheses of Photoreceptor Degeneration and Death in RP}
It is crucial to identify the precise cause(s) of both rod and cone cell death, as it is the secondary wave of photoreceptor death in RP patients that results in total blindness \cite{Bhatt,Rivas}. Substantial experimental research (consisting mainly of animal models) has been devoted to this effort.
There are several hypotheses/theories as to what leads to photoreceptor death in RP, but none of them have been definitively proven \cite{Bhatt}.  All theories suggest indirectly or directly that mutations result in gene products that are damaging to photoreceptors (including their signaling pathways and cell-cell interactions) or that mutations cause lack of a protein that is needed for photoreceptor survival \cite{Shen}.  Among these theories is one which led to the development of antioxidant therapy for cone survival and is based on the hypothesis that the death of rods leads to higher oxidative stress in the retina, which eventually kills the cones \cite{Bhatt}.  This was demonstrated by showing that antioxidants promote cone survival and vision in rd1 mice as they get older. Another theory comes from the idea that the rods produce some kind of signal that is essential for maintaining cone viability, so that the disappearance of rods for whatever reason would deprive the cones of this signal and trigger their degeneration \cite{MohanSaid}.   This theory was supported through various experiments on mice and rats and led to the discovery of the RdCVF.  A third theory comes from the synergistic interactions of the RPE and the photoreceptors since defects in genes normally expressed in the photoreceptors can lead to degeneration of the RPE and vice versa \cite{Strauss}.  The inability of the RPE to secrete neuroprotective and growth factors such as PEDF, bFGF, and CNTF that can protect photoreceptors from damage has been implicated in the pathogenesis of retinal diseases (\cite{Strauss} and references therein). Inappropriate signaling between photoreceptors and bipolar cells whose signaling pathway remains intact for RP patients has also been postulated as a potential reason for the secondary death wave of photoreceptor in rod-cone RP patients \cite{John}. Toxicity released by mutated and degenerate rods is another hypothesis but this theory has been challenged by experiments which showed that cone death is uniform in space irrespective of where they are in relationship to the mutated rods \cite{sa,MohanSaid}. Another related hypothesis is that microglia activated by ongoing rod death migrate to the photoreceptor layer and secrete toxic substances that kill cones \cite{Gupta}.

Depletion of photoreceptors or abnormally short OS that result in oxygen toxicity only explain part of the cause of cell death in RP \cite{Travis,Stone}. Although rods die by apoptosis, which generally avoids release of potentially toxic cellular metabolites and prevents extension of cellular breakdown to surrounding regions of tissues, it seems possible that degeneration of the abundant rods (numbering about 90-120 million in the human eye) could have adverse effects upon the adjacent cones and nearby bipolar cells \cite{MohanSaid}.  Even though there are some modifications in the inner retinal neurons these nearby neurons do not undergo a widespread death following the first wave of photoreceptors death, as one would expect.  The main flaw in these toxicity theories is that cone cell death occurs very slowly and continues long after all rods have died. While toxicity might explain photoreceptors degeneration in the short term (at the early stage of the disease) it fails to hold in the long run \cite{MohanSaid}.

The theory of oxidative stress fails to completely explain the sequential photoreceptor death in reverse RP (or in cases when both cones and rods die simultaneously) and in the rod-cone RP cases it raises the questions as to why rods die preferentially and why only cones are saved when treated with antioxidants.  In an attempt to address this researchers have postulated that rods might be more sensitive to reactive oxygen species (ROS) and thus they die first, which then probably further increases the oxidative stress in the retina. The generation of ROS, by proteins in various retinal layers, is part of the normal maintenance of the retina \cite{Bhatt}. Rods are more numerous than cones and are metabolically active cells with a high level of oxygen consumption. Choroidal vessels are not subject to autoregulation by tissue oxygen levels and as death of rods occurs, the tissue level of oxygen in the retina increases \cite{Yu2000,Yu2004}.  Oxygen levels in the outer retina vary inversely with oxygen consumption; as the number of rods in a particular area of the retina decreases, oxygen consumption decreases, and thus oxygen levels increase. Based on this, one would expect regions of human retinas with the highest rod/cone ratio (the far periphery of the eye) to be the first regions to experience photoreceptor death due to their exposure to dangerously high levels of oxygen \cite{Shen}.  This is what we see in the case of rod-cone RP where the peripheral visual field is constricted first but not in the case of reverse RP. While this theory might not depict the entire picture it provides one of potentially multiple causes of cell death in RP as it has been established experimentally that high tissue levels of oxygen are damaging to photoreceptors and result in cell death \cite{Yamada,Okoye}.

Another valuable insight for the death of rods and cones in RP comes from our current understanding of the RPE as a vital component of the visual system that supports normal photoreceptor function.  Integrated interactions between photoreceptors and the RPE are key in the visual process.  Among these interactions are the visual cycle (in which the structure of retinal is changed by light transduction: in photoreceptors from 11-cis- to all- trans-retinal and in the RPE from all-trans- to 11-cis-retinal) and the phagocytosis of photoreceptor OS (which contributes to the photoreceptors shedding process and the optimal photoreceptor structure necessary for light perception).  Experiments have revealed that signs of photoreceptor apoptosis due to intensified light exposure were fully reversible potentially due to increased secretion and synthesis of different neuro protective factors by the RPE such as PEDF, CNTF, and bFGF, which are known to be protective against light damage \cite{Strauss}.  Among the essential proteins linking the RPE and photoreceptors is the interphotoreceptor retinoid binding protein (IRBP) which is primarily a product of the rods, with less of a contribution from the cones \cite{Hollyfield,Rodrigues,Porello,MohanSaid}.  A decrease in IRBP due to the death of rods as well as a loss of the RPE-recycled products of rod OS shedding may contribute to the observed cone OS shortening and ultimately, the cone cell death found in RP retinas.  This is mainly attributed to an associated local deficiency in vitamin A, as it is known that vitamin A deficiency causes photoreceptor OS shortening and cell death \cite{Dowling,Carter}.  
The inability of the RPE to provide essential neuro protective and growth factors to individuals with RP opens a realm of possibilities that call for a thorough investigation of the interactions of photoreceptors and their exchange of trophic factors.

Another protein that seems to be key in the interactions of the photoreceptors responsible for the viability of the cones is the RdCVF.  RdCVF is a truncated thioredoxin-like protein specifically expressed by the rods.  Experiments demonstrated a 40\% rescue effect in cones and thus slowed down the degeneration of cones in chick and mouse models \cite{Leveillard,Yang}. Additional experiments revealed that RdCVF had a bigger effect in the functionality of cones than in their viability \cite{Yang}.  Although RdCVF is a good candidate to target for cone survival in RP, a number of issues remain to be addressed. For example, the mechanism of action underlying the cone survival effect needs to be better understood as well as the intracellular signaling pathways that are involved \cite{Sahel}.  Another issue that is currently being researched is the delivery mechanism of RdCVF as injection of a protein but this might not to be a satisfactory option for the long term. Gene-based delivery through viral vectors or using encapsulated cell technology would ensure a more stable and persistent release of RdCVF \cite{Sahel}.
Irrespective of the mutation, the final common pathway of the first wave of photoreceptor death is degeneration through apoptosis; the mechanism underlying secondary wave of cone (in rod-cone RP) or rod (reverse RP) death remains unknown \cite{MohanSaid}.

\section*{Mathematical Model}
\begin{figure*}
\includegraphics[width=6 in, height=6 in]{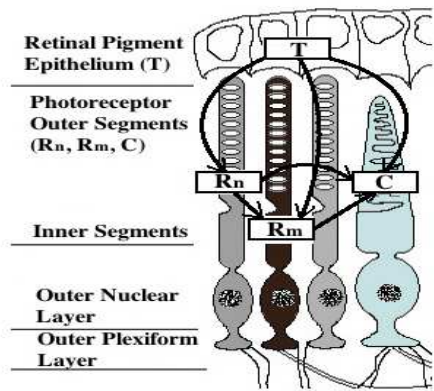}
\caption{ Flow-diagram of photoreceptor interactions affecting RP.  Here we assume two distinct classes of rod cells depending on whether or not they have had their functionality compromised in any way as a result of the genetic mutation and distinguish them as normal and mutated rods, $R_n$ and $R_m$ respectively. For these different types of rods, the rate at which they draw from and utilize the trophic factors and their contribution of Rod-derived Cone Viability Factor (RdCVF) may be different\cite{Leveillard,MohandSaid1998,Wenzel,Leveillard2010}.  In this work, we do not attempt to distinguish between the causal mutations of RP.  $T$ represents the RPE cells, which mediate the passage of growth factors, metabolites, ions, etc.~to the photoreceptors.  Fractions of $R_n$, $R_m$, and $C$ (cones) are interpreted as the outer segment (OS) discs whereas fractions of $T$ are viewed as the nutrients, etc.~that are brought to the photoreceptors via the RPE\cite{camacho2,Strauss}. }\label{FlowChart}
\end{figure*}

In an attempt to shed light into the underlying mechanisms and factors involved in the development of the different types of RP we create a mathematical model of photoreceptor interactions that explicitly incorporates RdCVF and implicitly includes other essential photoreceptor neuroprotective and growth factors secreted by the RPE. 
We consider two different classes of rod cells, mutated and normal, to represent those that have had their functionality affected by the genetic mutation(s) and those that have not\cite{camacho2,CamachoThesis,Burns} (see Figure \ref{FlowChart} for a simple physiological illustration).  This distinction between the two classes of rods is essential due to the numerous genes and processes affected by the mutations associated with RP\cite{Phelan,Shintani,Shen} and the great variability of the disease\cite{Tsujikawa}.  While much RP research has characterized the different types of mutations that cause RP, we are not modeling and do not address how the mutation occurred or what type of mutation it is or what it affects. Rather, our model focuses on the interaction and structure of the photoreceptors and their OS in relation to the renewal and shedding process.  We simply assume that there is a mutation that has already been expressed and therefore some of the rods cells can be thought of as ``mutated rods'' \cite{Tsujikawa}. We similarly do not model the visual transduction or retinal metabolism\cite{Keener}.   Thus our model can be thought of as representative of the photoreceptor and RPE interactions at or after the onset of RP\cite{camacho2}.
As is often done, we used previous mathematical models of the eye together with experimental evidence to assume that both eyes are identical and thus consider only one eye \cite{CamachoThesis,camacho2,CamachoRR,Shintani}. This work builds on our previous mathematical model of photoreceptor interactions in a healthy eye but now we incorporate mutated rods that are present in RP.  For the sake of mathematical simplicity, we also ignore any spatial
effects, for example, the different spatial distribution of the photoreceptors at various locations in the retina \cite{Burns}.   Two other mechanistic models of RP focus on the kinetics of neuron degeneration and lateral inhibition of nearby cells and not on cone and rod photoreceptor interaction\cite{Burns,Tayyab}.  

In order to focus on the mechanisms behind the various manifestations of RP, we create a mathematical model that includes two different classes of rod cells, $R_m$ and $R_n$ (corresponding to mutated and normal rod OS, respectively), one type of cone OS (green, red, blue cone OS are grouped together), $C$, and a trophic pool (representative of the RPE's neuroprotective and growth factors) from which the rods and cones draw their nutrients, $T$.
We utilize the framework from epidemiology and population dynamics to model the interactions between the photoreceptors and trophic factors as a predator-prey system; see Figure \ref{FlowChart}. Even though no new rods or cones are created in a mature retina, the process of shedding and renewal lends itself to the
interpretation of birth and death of the rods and cones and we incorporate this interpretation into our model \cite{Bok,Hendrickson,Papermaster}.  We view our units as ``cells'' and therefore fractions of a cell are the POS in these three types of photoreceptors. Experiments have shown that in a rod-free retina the absence of cone OS corresponds to blindness even though the living cone cell remains\cite{Santos, Guerin}.  We will take this same perspective. Similarly the nutrients represent fractions of the trophic pool consisting of growth factors, metabolites, ions, and water (e.g., PEDF, CNTF, bFGF, Gas6, PDGF, ATP, glucose, retinal, and omega-3 fatty acids) to and from the photoreceptors provided or secreted by the RPE \cite{Strauss,Murakami,Longbottom,LaVail1998,Li,Wenzel,Frasson}.
In RP, degeneration can also extend to the outer retina nuclear layers and the ganglion cell layer but we focus this work only on the photoreceptor cells \cite{Santos,Flannery}. Our model extends the healthy eye model of \cite{camacho2}: 
\begin{eqnarray}
\dot{R_n}&=&R_n(a_n T- \mu_n-m) \nonumber \\
\dot{R_m}&=&R_m(a_m T- \mu_m)+mR_n \nonumber \\
\dot{C}&=&C(a_cT -\mu_c+d_nR_n+d_mR_m) \nonumber \\
\dot{T}&=&T(\Gamma-kT-\beta_n R_n-\beta_m R_m-\gamma C),
\label{full1}
\end{eqnarray}
where all parameters are nonnegative; see Table \ref{Table2} for definitions of  the parameters.  We do not distinguish between the various factors that constitute the trophic factors so as to focus on the critical interactions of the rods and the cones\cite{camacho2}.  Furthermore, this allows our work to stay consistent with all experiments in the literature where no distinction between a rod-trophic and a cone-trophic pool is made \cite{Leveillard,Bhatt}.  Our model assumes that both the rods and the cones take more nutrients from the trophic pool than they contribute (as recycled products) and that they convert
these trophic factors to new OS discs at some ratio. 
To the best of our knowledge, no experimental work has been done to date that discusses the production rate of trophic factors in the {\it absence} of photoreceptors; this current work supposes that the production of the trophic factors is logistic in nature and thus will reach an equilibrium (governed by $\Gamma$ and $k$) in the absence of photoreceptors, instead of growing without bound\cite{camacho2}. The structure of this mathematical model incorporates the presence, through the parameters $d_m$ and $d_n$, of the RdCVF protein that helps cone survivability and functionality yet does not negatively influence the rod population \cite{Leveillard,MohanSaid,Shen,Sahel}.  Experimental evidence suggests that the RdCVF directly affects the cones and thus we model this as a direct rod-cone interaction, the necessity of which was already established in \cite{camacho2}.  The literature does not suggest that mutated rods stop or decrease production of RdCVF and thus our model includes this possible contribution of RdCVF by the mutated rods via $d_m$.  This model reduces to the previous work in \cite{camacho2} if we ignore the terms related to the mutated rods, i.e., $m=R_m=d_m-\beta_m=0$, and do not consider saturation of the trophic pool, $k=0$.

We define the following terms:
\begin{equation}
D_m=\frac{\mu_m}{a_m}, \quad D_n=\frac{\mu_n+m}{a_n}, \quad D_c=\frac{\mu_c}{a_c}, \quad D_T=\frac{\Gamma}{k},\label{D_terms}
\end{equation}
which are key in our analysis.  Based on the equations (\ref{full1}), we determine steady state solutions $(R_n,R_m,C,T)$ that correspond to the various physiological stages of RP and whose corresponding stage and subtype of RP is summarized in Table \ref{Table1}.  Our system yields seven equilibria given by

\noindent
\begin{eqnarray}
E_1&:& (0,0,0,0), \nonumber \\ 
E_2&:& (0,0,0,D_T), \nonumber \\ 
E_3&:& \left(0,\frac{B_1}{a_m\beta_m},0,D_m\right), \nonumber \\ 
E_4&:& \left(0,0,\frac{B_2}{a_c\gamma},D_c\right), \nonumber \\ 
E_5&:& \left(0,\frac{B_3}{a_md_m},C_5^*,D_m\right), \mbox{\quad where } C_5^*=\frac{d_m B_1 -\beta_m B_3}{\gamma a_m d_m} \nonumber \\
E_6&:& \left(R_{n6}^*,R_{m6}^*,0,D_n\right), \mbox{\quad where }\nonumber \\
&& \qquad R_{n6}^*=\frac{A_1A_3}{a_n(m\beta_m a_n+\beta_n A_1)}, 
R_{m6}^*= \frac{mA_3}{m\beta_m a_n+\beta_n A_1}, \nonumber \\
E_7&:& \left(R_{n7}^*,R_{m7}^*,C_7^*,D_n\right), \mbox{\quad where } \nonumber \\
&& \qquad R_{n7}^*=\frac{A_1A_2 }{a_n(d_n A_1+d_m a_n m)}, 
R_{m7}^* = \frac{mA_2}{d_n A_1+d_m a_n m}, \nonumber \\
&&\mbox{ and } C_7^*= \frac{1}{\gamma a_n}\left[A_3-A_2\left(\frac{m\beta_m a_n+\beta_n A_1}{d_n A_1+d_m a_n m}\right)\right],\label{equilAll}
\end{eqnarray}
with\\
\makebox[2.5in][l]{$\qquad$ $A_1= a_na_m\left(D_m-D_n\right)$,}
$A_2= a_na_c\left(D_c-D_n\right)$,  \\
\makebox[2.5in][l]{$\qquad$ $A_3= a_nk\left(D_T-D_n\right)$,}
$B_1= a_mk\left(D_T-D_m\right)$,  \\
\makebox[2.5in][l]{$\qquad$ $B_2= a_ck\left(D_T-D_c\right)$,}
$B_3= a_ma_c\left(D_c-D_m\right)$.

The ratios defined by the $D_i$ for $i=m,n,c,T$ represent key quantities that govern the progression of RP and will be discussed in Appendix 1.

\section*{Progression of RP through Equilibrium Solutions}
Analysis of the seven steady state solutions (\ref{equilAll}) give conditions on the parameters for the linear stability of the various stages of RP \cite{MohandSaid1998}.  Once a particular stage of RP has been reached, these conditions represent quantities that must persist in an individual to prevent further progression of RP (e.g., moving to a more advanced stage of RP) and equivalently they give the parameters involved if RP progression continues.   These conditions involve the terms in (\ref{D_terms}): $D_m$, $D_n$, $D_c$, and $D_T$, which turn out to be the key quantities predicted by our model to govern the progression of RP.\footnote{A non-dimensionalization of the original system, while reducing the parameters from 14 to 10 yields the same 4 essential ratios in (\ref{D_terms}) that are involved in the various bifurcation sequences, without additional insight gained to date with the non-dimensional system.  Thus, in order to keep the problem tied to the physiology as much as possible, we leave our system in its original dimensional form and focus on the relationship between the $D_i$ as the mechanisms governing the progression of RP.}  The first three of these terms are the ratios of the OS departure (due to shedding and/or mutation) to the renewal for a given class of photoreceptors.  The fourth term, $D_T$, is the ratio of the natural growth rate to the carrying capacity of the trophic factors in the absence of any photoreceptors (complete blindness).  Our results are supported by previous experimental research suggesting that disturbances or imbalances of disc shedding to disc formation is associated with RP \cite{Ripps1978}.  

Linear stability analysis of the steady state solutions $E_i$, $i=1,\cdots, 7$ allow us to describe the progression of RP for various subtypes within rod-cone RP and reverse RP.\\

\noindent {\it Result:} {\it We have observed the equilibria of system (\ref{full1}) in multiple transcritical bifurcations yielding the following progressions:
\begin{enumerate}
\item[1.] $E_7\to E_4\to E_2$; $E_7\to E_5\to E_4\to E_2$ (rod-cone RP)
\item[2.] $E_7\to E_6\to E_2$; $E_7\to E_6\to E_3\to E_2$  (cone-rod RP)
\item[3.] $E_7\to E_5\to E_3\to E_2$ (cone-rod RP or simultaneous photoreceptor death)
\item[4.] $E_7\to E_2$;  (simultaneous photoreceptor death)
\end{enumerate}
In each of 1-4, a given equilibria $E_i$ is stable and loses its stability through a transcritical bifurcation ($\to$) as the subsequent solution becomes stable. The stability of the equilibria depends on $D_m,D_n,D_c,D_T$. }\\

While not all of the transcritical bifurcations are obvious from inspection, calculations in a computer software package such as Maple show that at each bifurcation, the two equilibria become one as they switch (at least) one direction of stability for realistic parameter values.  When $D_n=D_m=D_c=D_T$ in parameter space, equilibria $E_2, \cdots, E_7$ coalesce in an equilibrium point with three zero eigenvalues.  We have examined the flow in the corresponding three-dimensional center manifold of the non-dimensional version of (\ref{full1}) but have not gained any additional insight regarding the observed bifurcations.  The numerical results on the center manifold flow have confirmed the numerical results in the original system regarding the stability of only one equilibrium being relevant and stable at any given point in the realistic range in parameter space.  However, we have observed stable periodic solutions for certain (unrealistic) parameter values.  Following these solutions with MatCont has not yet allowed us to conclude the bifurcation that gave rise to these limit cycles (one exists in the full 4-dimensional space while the other exists in the 3-dimensional subspace given by $R_n=0$).  This aspect of the work is still ongoing.  Thus, while we have analytically observed the sequence of bifurcations in the above {\it Result} for specific realistic parameter values, we are only able to conjecture that this result holds for all choices of parameter values.  The statements below are thus certain for realistic parameter values and conjectured to hold in general.  See Appendix 1 for a discussion of the analytical results regarding stability of the equilibria, Appendix 2 for a discussion about parameter estimation of the model, and Appendix 3 for other stable solutions observed.

It is important to observe that $E_4$, for example, represents a state in which both types of rods have died off and only cones remain; however, the number of cones present in a given $E_4$ value may differ widely and depends on the parameters.  Thus, the bifurcation $E_7\to E_4$ allows for the possibility that the cone levels in $E_7$ have remained the same as the rods die off (as the values in $E_7$ get closer to those in $E_4$ as the parameters are varied), but also allows for the possibility that some/many of the cones have also begun to die off.  Thus while we can think of $E_4$ as the typical state in rod-cone RP in which all rods have died off while the cones remain, we must remember that the level of cones in $E_4$ may not be the original numbers of cones seen in a healthy eye.  Keeping this in mind, we can interpret the sequences of transcritical bifurcations of the {\it Result} as various subtypes of RP by asking ``which photoreceptors remain just after the bifurcation'' as we vary a given parameter or set of parameters.

The sequences given in 1 describe the progression of rod-cone RP via two different paths.  The transitions $E_7\to E_4\to E_2$ might represent a more aggressive subtype of RP in which night blindness occurs because of a simultaneous loss of mutated and normal rods ($E_7\to E_4$) before the complete death of cones ($E_4\to E_2$).  The sequence $E_7\to E_5\to E_4\to E_2$ might be a less aggressive subtype as the normal rods die off first ($E_7\to E_5$) followed by the mutated rods ($E_5\to E_4$) before the cones die off. In both sequences, we have an intermediate state in which only cones remain before eventually dying off.
The sequences given in 2 describe the progression of cone-rod RP via two different paths, with the latter having the interpretation of both cone-rod RP and simultaneous death of rods and cones depending on the actual equilibrium values. The transition $E_7\to E_6\to E_2$ might represent a more aggressive form in which cones die off before either normal or mutated rods die off ($E_7\to E_6$) and then both types of rods die off together ($E_6\to E_2$). 
In the path $E_7\to E_6\to E_3\to E_2$, the cones are predicted to die off first  ($E_7\to E_6$), then the normal rods ($E_6\to E_3$), followed by the death of the mutated rods rendering complete blindness ($E_3\to E_2$).
The sequence $E_7\to E_5\to E_3\to E_2$ (in 3)  might represent a somewhat less aggressive form of cone-rod or simultaneous death RP in which only normal rods die off ($E_7\to E_5$) before the cones completely disappear ($E_5\to E_3$).  In this sequence the individual will have limited night vision and thus clinically this would correspond to a cone-rod RP as long as night vision persists to some degree.  In these three sequences (of 2 and 3), it is a type of rod that is the last photoreceptor to die off thus leading to the classification of all three as cone-rod RP.  In contrast, the sequence given in 4, $E_7\to E_2$, describes an additional sub-type of RP in which cones and rods die off simultaneously.  If the functionality of the rods has been extremely affected by the mutations to the point where night vision has been completely impaired such that $E_3$ contains no functional rods at all then the sequence $E_7\to E_5 \to E_3\to E_2$ can be considered to be in category 3 where clinically cones and rods are considered to die off simultaneously.  For the manner in which RP is addressed in this work photoreceptors that contained OS that shed and renew are considered to be functional to some degree.

It is important to again note that some of the above sequences, for example, $E_7\to E_4\to E_2$ may in some situations be clinically classify as the same subtype as $E_7\to E_2$ if the cone levels of the $E_4$ equilibrium are very close to zero and disappear very shortly after the rods disappear. To the best of our knowledge, no experiments have made the distinction between normal and mutated rod death during the progression of any sub-type of RP nor has any research been able to follow the complete progression of death for a given type of photoreceptor (i.e., produce a time series for the cones or rods in a given mammalian subject).  The results predicted by this model thus provide a potentially crucial understanding of the progression of RP in terms of the parameters that give rise to these stage changes.  This knowledge gives the framework that will guide future experiments focusing on potential therapies to slow the inevitable progression of RP.  Insights into what must be considered comes from the stability conditions of the transcritical bifurcations.

Since the terms $D_n, D_m, D_c, D_T$ are all involved in the conditions for the stability and/or physiological existence of equilibria $E_2,\cdots,E_7$, future experiments exploring potential long term therapies and interventions would need to keep in mind these quantities and their interrelatedness as these yield the  multi-faceted therapies that can halt RP brought about by the change in these quantities. By varying between one and three parameters of the system (\ref{full1}), we can thus observe many different progressions of cone-rod RP and rod-cone RP (see Figures \ref{Bifn1} and \ref{Bifn2}).  Our model suggests examining experimentally the causes of the shedding and renewal imbalances and of changes in the trophic pool, mathematically seen as changes in $D_m$, $D_n$, $D_c$, and $D_T$, and studying how to alter these quantities in RP patients \cite{Ripps1978}.  In our system, we have observed exactly one stable equilibrium $E_i$, $i=2,\cdots, 7$ that is stable for a given choice of realistic parameters.  This limits the possibilities for individuals with RP and thus it becomes even more crucial to have a better understanding of how to clinically control the ratios $D_m$, $D_n$, $D_c$, and $D_T$.

\section*{Conclusion}
The work presented here offers the first quantitative model of photoreceptor interactions that can track the progression of RP through all the various stages and subtypes of this disease and thus offers alternatives to stop its evolution at each stage. 
Our model predicts the progression of rod-cone RP ($E_7\to E_4\to E_2$ and $E_7\to E_5\to E_4\to E_2$) and cone-rod RP ($E_7\to E_6\to E_2$, $E_7\to E_6\to E_3\to E_2$, and $E_7\to E_5\to E_3\to E_2$) by examining the ratios in the quantities $D_m$, $D_n$, $D_c$, and $D_T$. Because the stability of the equilibria are governed by these ratios, our model suggests that future physiological experiments and research should focus on the shedding and renewal for each class of photoreceptors as well as the carrying capacity of the trophic factors as potential leads into long-term treatments for RP \cite{Ripps1978}.  

\begin{figure*}
\begin{tabular}{p{2.8in}@{}p{2.8in}}
\resizebox{2.8in}{2.5in}
{\includegraphics{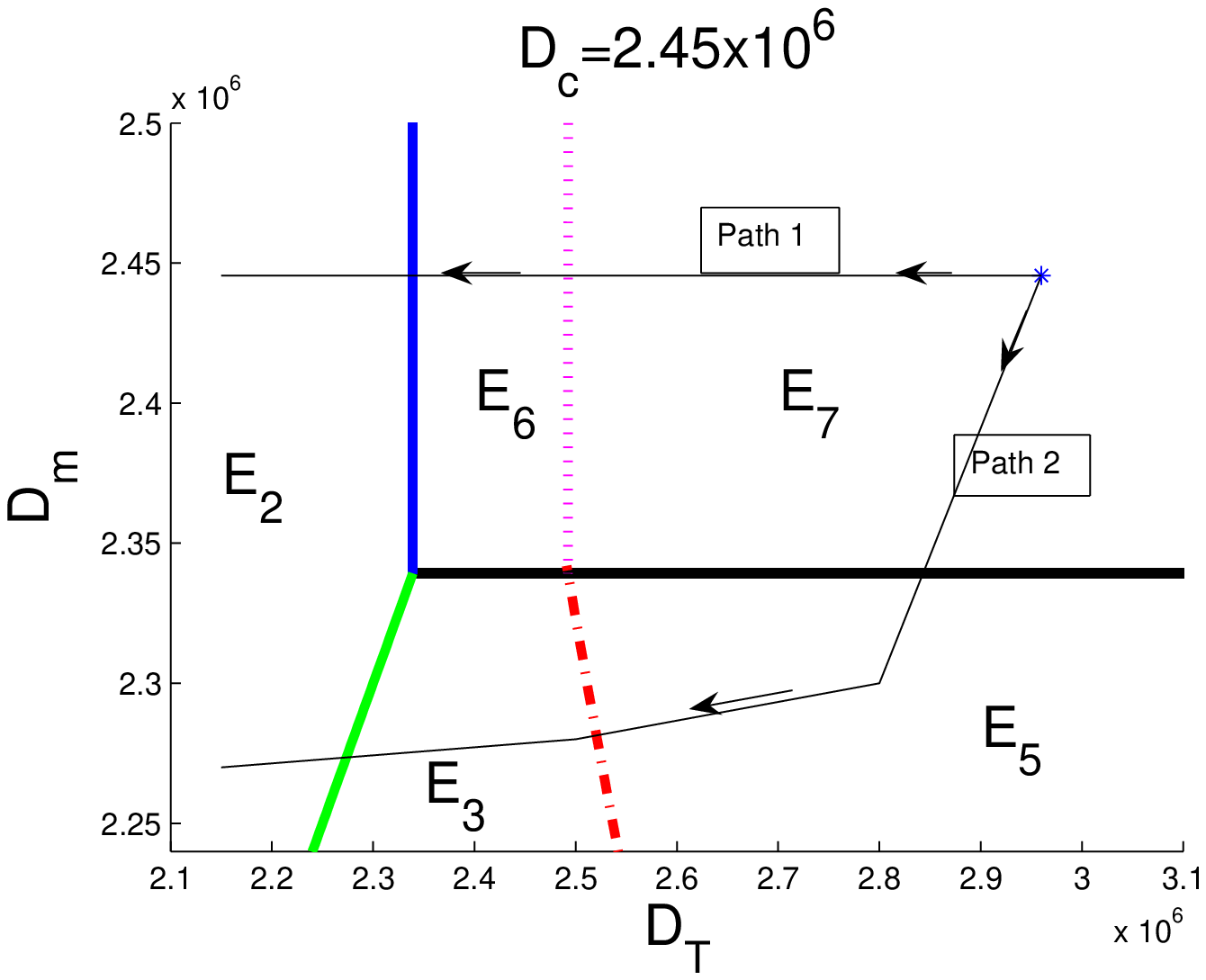}} & 
\resizebox{2.8in}{2.5in}
{\includegraphics{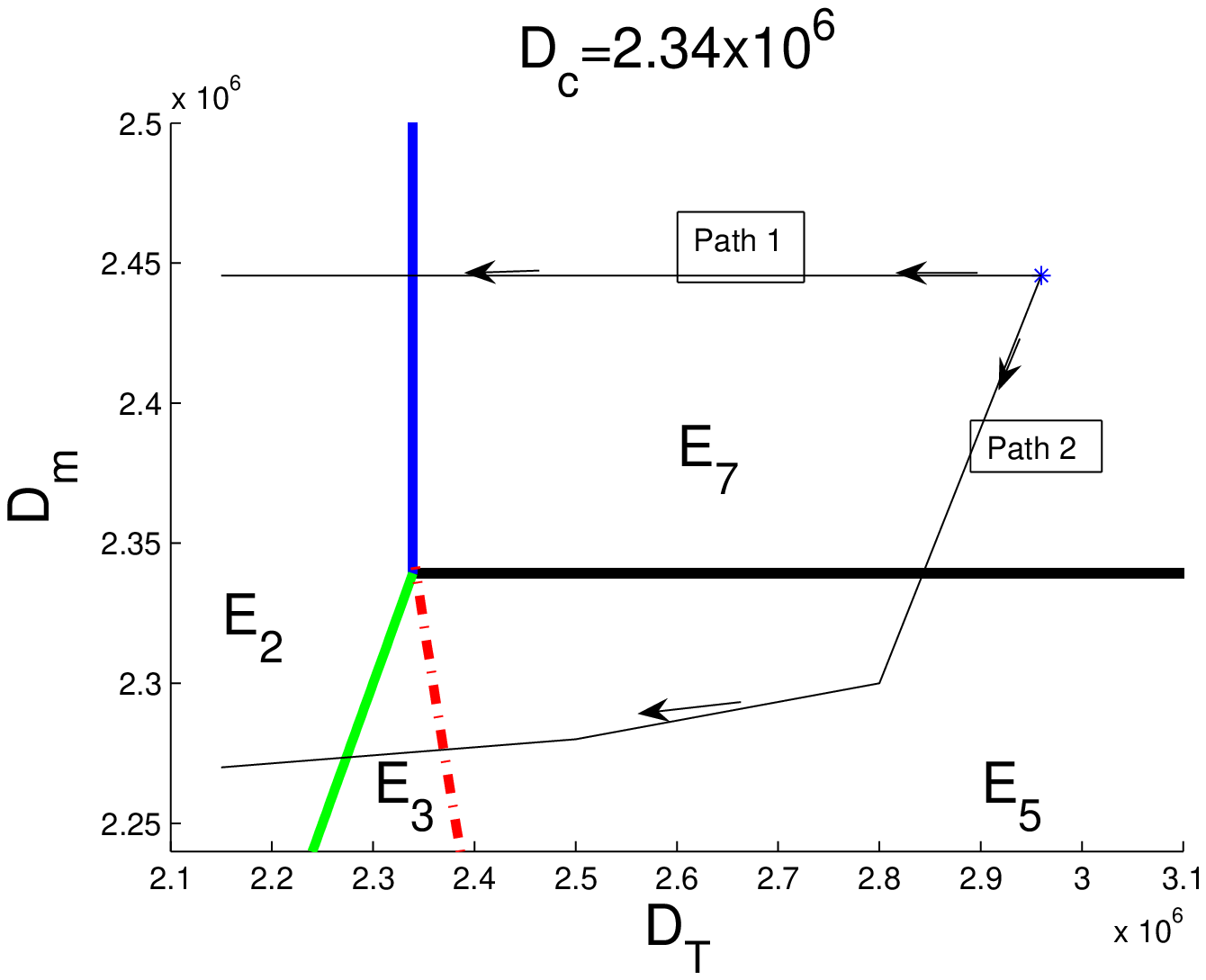}} 
\end{tabular}
\caption{ {\bf Possible paths to blindness in two hypothetical patients with cone-rod RP.} The planes illustrate the regions where various stages of RP are present in two identical patients (i.e., the same $D_n$, $\gamma$, etc.) except for certain aspects of their cone shedding to renewal ratio, $D_c$. Both $D_c$ values (as well as all other parameter values) are within the range of realistic values.  The blue asterisk represents the parameter values from the best fit (lowest MLE).  The solid dark lines that separate the regions of parameter space correspond to transcritical bifurcations of the $E_i$.  Sample paths 1 and 2 in both patients trace the same two sets of values for $D_m$ and $D_T$. Path 1 in patient 1 ($D_c=2.45\times 10^6$ and is along $D_m\approx 2.45\times 10^6$) illustrates the path $E_7$ (all photoreceptors present) $\to E_6 $ (cones die off) $\to E_2$ (all rods die off resulting in complete blindness)  that corresponds to a cone-rod RP path.  The same sample path in patient 2 $D_c=2.34\times 10^6$  is represented by $E_7\to E_2$  and corresponds to simultaneous rod-cone death. (When varying only two parameters there is a very narrow region of $E_6$ stability, suggesting that $C$ dies out first; when varying three parameters, we can obtain $E_7\to E_2$.)  The steady state solutions $E_i$ undergo transcritical bifurcations and change continuously throughout the plane so that near a bifurcation set (represented by the bold lines/curves) two steady state solutions are ``close'' in value.  For example as we follow the path $E_7\to E_2$ the cones and both types of rods decrease continuously as the parameter changes and just before crossing the bifurcation line/curve into region $E_2$ all photoreceptors are nearly zero, corresponding to a patient being almost blind.  Sample path 2 corresponds to the same progression in both patients; $E_7$ (all photoreceptors present) $\to E_5$ (normal rods die off ) $\to E_3 $ (cones die off) $\to E_2$ (mutated rods die off giving total blindness) but the range of parameter values in each stage (assuming identical reduction in trophic factors and in shedding to renewal ratios) is different.} \label{Bifn1}
\end{figure*}

The stability of the various stages of RP depend on the difference of these ratios, $D_i-D_j$, for $i\neq j$ and $i,j=m,n,c,T$ given above.  The steady state solutions, representative of the RP stages, become physiologically relevant through transcritical bifurcations, when two steady states coincide and then switch their stability\cite{eoa}. Analytical and numerical evidence indicates that only one of the steady state solutions is stable for realistic parameter values; for certain unrealistic parameter values, bi-stability has been observed (see Appendix 3).  By varying two parameters, we can track the progression of RP as we cross the regions in which a given steady state solution representing the current stage of RP becomes unstable; see Figure \ref{Bifn1}.

Progression of RP culminating in complete blindness means we begin in the region where $E_7$ is the stable solution (first stage of RP) and end up in the region where $E_2$ is the stable solution (complete blindness). Sample paths 1 and 2 in Figure \ref{Bifn1} illustrate the variability in RP and how identical changes in the parameters in two hypothetical different patients (both with realistic parameter values) could yield different clinical manifestations and even sub-types of RP\cite{Hamel,Shen,Wong}.  In path 1 we track the progression by varying the same $D_m$, $D_n$, and $D_T$ values of two hypothetical individuals with all other parameters fixed at the same values except different $D_c$ values ($2.34\times 10^6$ and $2.45\times 10^6$, respectively with both realistic) and observed very different manifestations of RP, $E_7\to E_6\to E_2$ versus $E_7\to E_2$. In sample path 2 we see that while the observed sequence of RP stages might be the same, the range of parameter values in each RP stage may vary drastically in two patients\cite{Weleber2005,Cepko2005} even when only one parameter is different in their visual system. The different ranges are indicative of the different options of available to halt the disease. 
This finding is consistent with the established variability of RP between individuals \cite{Shen,Cepko2005,Wenzel}.  Furthermore, this result opens the opportunity for multiple individual targeted therapies to slow the progression of RP.  

\begin{figure*}
\begin{tabular}{p{2.8in}@{}p{2.8in}}
\resizebox{2.8in}{2.5in}
{\includegraphics{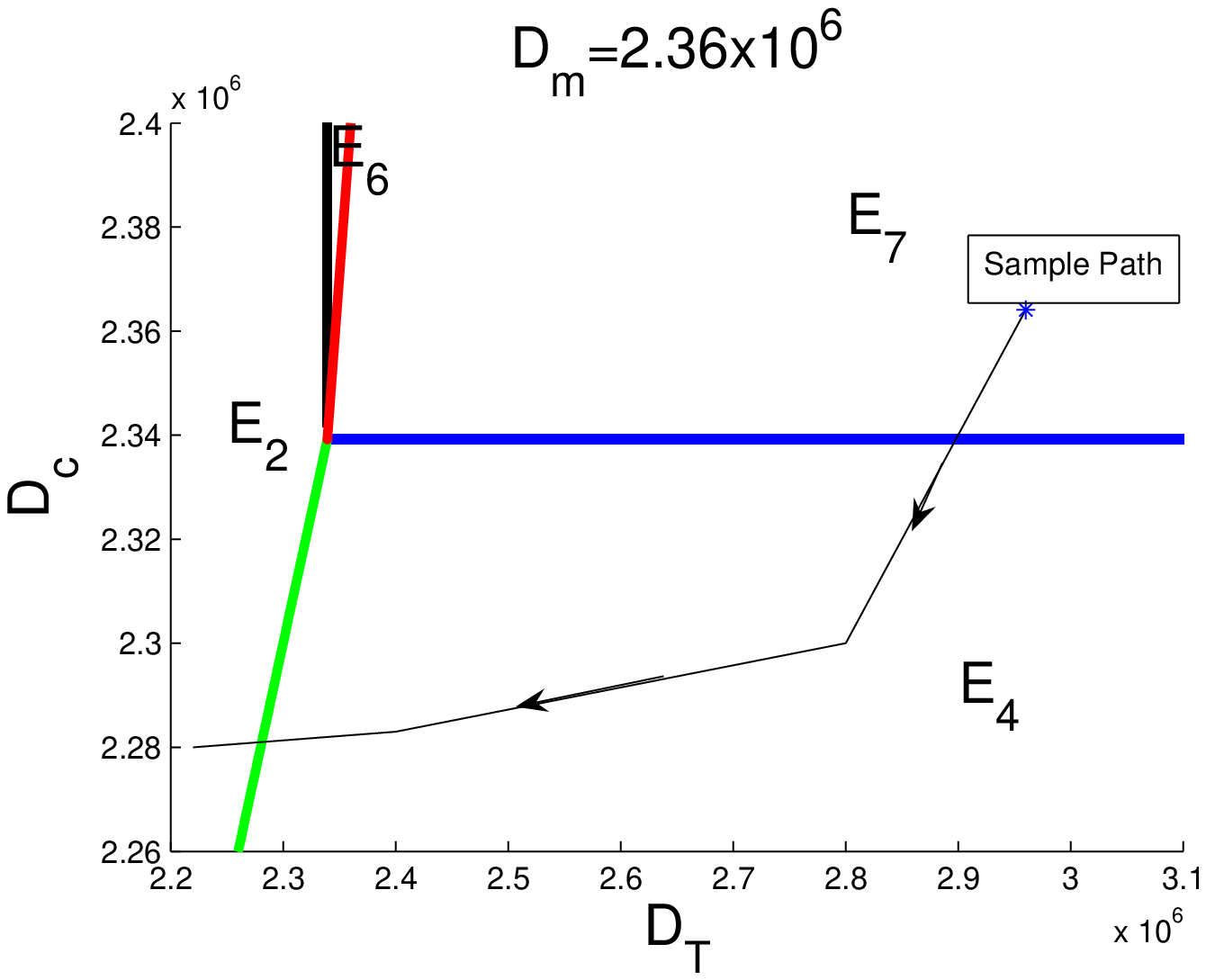}} & 
\resizebox{2.8in}{2.5in}
{\includegraphics{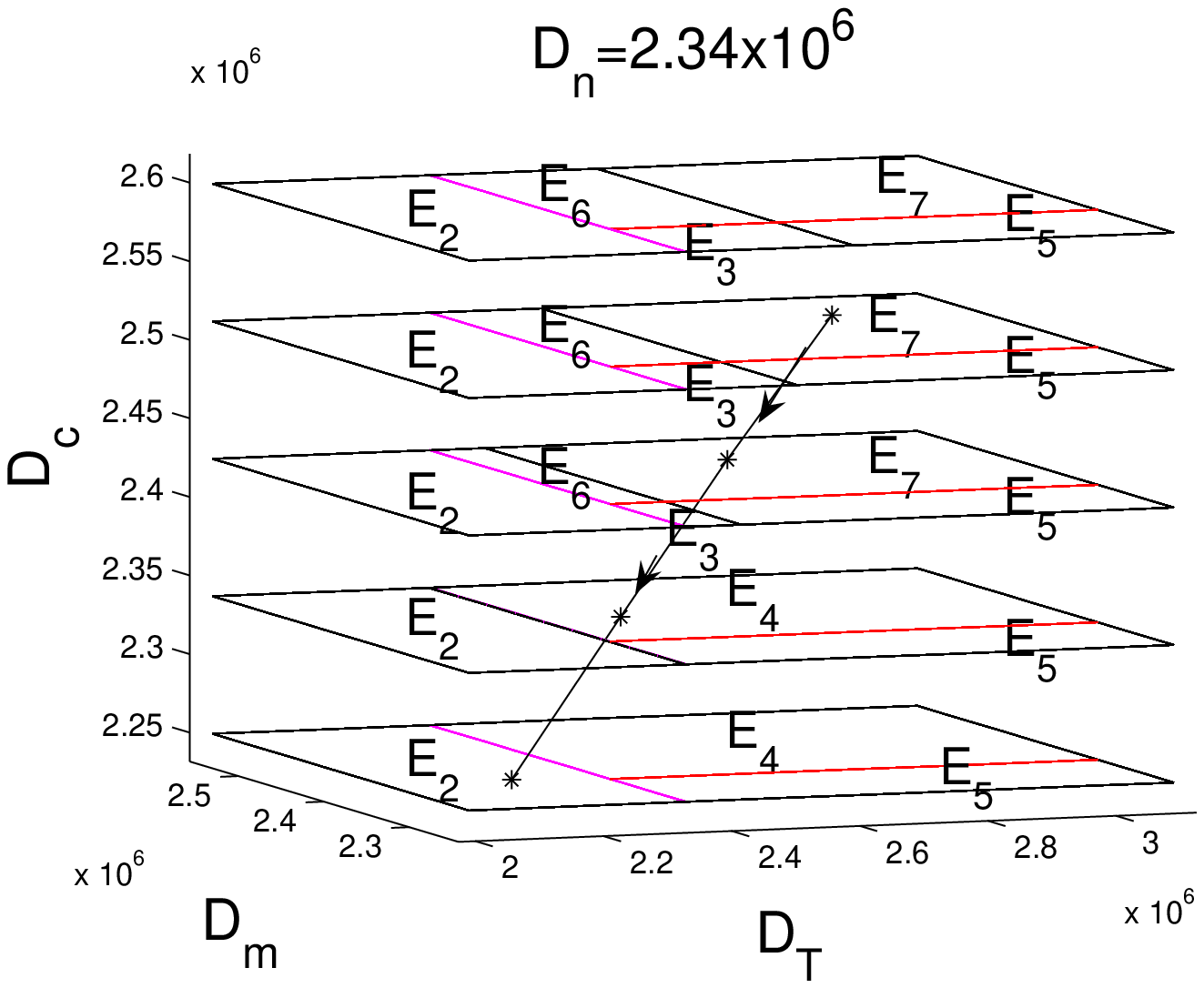}} 
\end{tabular}
\caption{ {\bf Possible paths to blindness in a patient with rod-cone RP.} In contrast to Figure \ref{Bifn1}, the left picture fixes the ratio $D_m$ and instead varies $D_c$ and $D_T$ (with $D_n$ the same in both pictures and all parameter values within their realistic ranges).  The solid dark lines that separate the regions of parameter space correspond to transcritical bifurcations of the $E_i$.   A particular progress of the common rod-cone RP is illustrated by the Sample Path. However, a more likely scenario is that three (or more) parameters might change in order to give the common sequence $E_7\to E_4\to E_2$.  In the right picture, we decrease the three ratios $D_m$, $D_c$, and $D_T$ to obtain the same clinically observed sequence of stages of RP. (The asterisk is at the intersection point of each plane and the numbers in the respective plane are the subscript of the stable steady state solution $E_i$.) At any given point, it will be the case that multiple options exist to slow the progress of RP by changing a combination of $D_m, D_c, D_T$ and thereby altering a given path in this parameter space. } \label{Bifn2}
\end{figure*}

While we may be able to see all possible stages of RP in an appropriate parameter region by varying at most two parameters, it is unlikely that this will be the norm physiologically since these biological components are all interconnected and disturbances in one can trickle down to others\cite{Wong}.  Thus Figure \ref{Bifn2} shows how we could progress from $E_7$ to $E_4$ to $E_2$ in the most common rod-cone RP by varying three parameters, thereby varying $D_m$, $D_c$, and $D_T$.  The relative size of the stability regions of $E_7$ and $E_4$ in this space illustrate the various ways to remain in the RP stages corresponding to $E_7$ or $E_4$ both of which permit day/color vision. The progression of rod-cone RP illustrated in the Sample Path (Figure \ref{Bifn2}) requires a decrease in the ratio of shedding to renewal of the cones ($D_c$) together with a decrease in the necessary trophic factors (i.e., nutrient starvation) for the vision system of this hypothetical individual to fail to meet the metabolic and functional needs of the photoreceptors (in the particular the cones). The changes in shedding, renewal, and trophic factors required to move from one stage to another depend on the differences of the relevant quantities $D_i-D_j$, $i\neq j$.  Counteracting these currently inevitable trajectories of rod-cone RP in our model translate to changing $D_c$ and such changes correspond to new therapeutic approaches that bring back a balance between shedding and renewal of the cone OS \cite{Ripps1978}. 

Currently there is no standard long-term treatment for RP, in part because the understanding of RP is incomplete.  There are many different theories for the physiological processes triggering photoreceptor death and RP progression.  Some of these causes, while still having physiological value, can be contradictory or can explain only certain aspects of RP progression, and/or only hold for the early stages of the disease.  Research in this area has improved the understanding of RP tremendously yet the possible mechanisms that trigger its inevitable progression and offer potential long term therapies are still open questions.
Our model provides a possible unified understanding to these two open questions. Recent experimental research has focused on nutritional and drug therapy, the visual transduction cascade, and the renewal and shedding of the POS \cite{Shintani,vanSoest,MohandSaid1998,Wenzel,Sahel,Fain,Strauss}.  For example, high dosage of vitamin A supplements have been utilized as a nutritional therapy as it has been shown to cause an increase in the trophic factors of nutrients essential for proper photoreceptor functioning\cite{Fain} and our model illustrates this and similar nutritional therapy in addition to growth factor delivery by an increase in $D_T$.  Likewise, changes in the shedding and/or renewal of the POS, which have been proven to be critical in the maintenance of healthy photoreceptors and are part of our main results, correspond to changes in $D_m$, $D_n$, or $D_c$ in our model depending on the specific type of photoreceptor that is affected.  Gene therapy, which is currently being explored experimentally, in our model corresponds to an impact on $D_n$ directly by replacing mutant or errant genes that affect normal rod shedding and/or renewal,
and perhaps the others indirectly if their renewal or shedding rates were affected  through changes in $D_n$ \cite{Leveillard2010}.

A promising new experimental approach to halt RP (and prevent total blindness) corresponds to maintaining the stability of $E_4$ (when only cones remain) in our model and it involves the discovery of the RdCVF\cite{Leveillard,Chalmel,MohandSaid1998,Wenzel,Leveillard2010}. Mathematically and physiologically, RdCVF is necessary in order for the cones' viability and thus injections of RdCVF offer a rescue effect on cone survivability and functionality in the absence of rods\cite{Yang}.  However, other approaches to avoid complete blindness arise from mathematically analyzing the stability of $E_4$ and the components that make it stable, which yield the conditions $D_c<D_n$, $D_T>D_c$, and $D_c<D_m$.  While $D_c$ is common in each and could be addressed by changes in the cone shedding or renewal, the other three quantities also must be considered to ensure that the cones-only steady state, $E_4$, remains viable.  A myriad of experimental treatments could be investigated to address these conditions.

RP is the term used to describe a collection of inherited degenerative eye diseases, which are not fully understood and for which reliable long term treatments still await\cite{Shintani,Phelan,MohanSaid,Hamel,MohandSaid2000,Murakami}.  While the vast majority of mutations have been identified in the genes that affect the rods, both rods and cones ultimately are affected\cite{Shen,Busskamp,Phelan,MohandSaid1998,Fain,Hamel}.  There are numerous other degenerative eye diseases that are not termed RP but that follow similar mechanisms of cell death triggered by mutations.  Given the general structure of our model examining photoreceptor interactions, we believe that this work will help spur mathematical and experimental research in this area with the analogous conclusions inferring multi-faceted approaches to slow the progression of many other degenerative diseases.

\section*{Acknowledgments}
We would like to thank the anonymous reviewers for their invaluable feedback on this work.  Part of this work was completed in the Mathematical and Theoretical Biology Institute (MTBI), which is supported with grants from the National Science Foundation, the National Security Agency, the Sloan Foundation, and the Office of the Provost of Arizona State University.

\section*{Appendix 1: Stability of Equilibrium Solutions}
In order to examine the stability of the steady state solutions (\ref{equilAll}), we consider the linear stability of each, together with the numerical observations (e.g., see Figures \ref{Bifn1}-\ref{Bifn2}):\\

\noindent {\bf $E_1= (0,0,0,0)$}\\
We can easily calculate the eigenvalues $\lambda_i=-(\mu_n+m), -\mu_m, -\mu_c, \Gamma$, which shows that $E_1$ is always a saddle point (unstable). \\

\noindent {\bf $E_2= (0,0,0,D_T)$}\\
This equilibrium is always physiologically relevant and denotes the trivial case in which there are no rods (of either type) and no cones but only trophic factors.  Physiologically this corresponds to total blindness in an individual.  The eigenvalues are again readily calculated as
$\lambda_i=\frac{1}{k}A_3$, $\frac{1}{k}B_1$, $\frac{1}{k}B_2$,
$-\Gamma$.  Any of the sub-types of RP could eventually lead to this state. \\

\noindent {\bf $E_3= (0,\frac{B_1}{a_m\beta_m},0,D_m)$}\\
The equilibrium $E_3$ represents the long term existence of mutated rods and trophic factors in the absence of cones and normal rods.  $E_3$ is physiologically relevant when 
$B_1>0$, which is exactly one of the conditions for $E_2$ to be unstable.  Thus as $E_3$ ceases to be physiologically relevant, it loses stability as $E_2$ gains its stability in a transcritical bifurcation, illustrated by the final stages of the sequences $E_7\to E_6\to E_3\to E_2$ and $E_7\to E_5\to E_3\to E_2$.
The eigenvalues of $E_3$ are given by
$$
\frac{A_1}{a_m}, \frac{d_m\gamma C_5^*}{\beta_m}, 
D_m\left(\frac{-k}{2}\pm \frac{1}{2}\sqrt{k^2-4B_1/D_m}\right)
$$
This will be stable for $A_1, C_5^*<0$ and $B_1>0$, the latter also being the condition for $E_3$ to be physiologically relevant.   \\

\noindent {\bf $E_4=(0,0,\frac{B_2}{a_c\gamma},D_c)$}\\
$E_4$ represents the situation where an individual only has day vision.  It illustrates the long term existence of the cones and trophic factors with both normal and mutated rods having disappeared.  It is a physiologically relevant solution when $B_2>0$, which is exactly one of the conditions for $E_2$ to be unstable. The eigenvalues are given by
$$
\frac{A_2}{a_c}, \frac{B_3}{a_c}, 
D_c\left(\frac{-k}{2}\pm \frac{1}{2}\sqrt{k^2-4B_2/D_c}\right).
$$
This equilibrium point corresponds to a stage in the typical rod-cone RP in which both types of rods die off before we observe the secondary wave of cone deaths (months and even years after all rods have disappeared)\cite{Shen}. Thus, we would expect this equilibrium point to be a temporary state in an RP patient that physiologically ceases to exist when the natural carrying capacity, $k$, of the trophic pool falls below a threshold value.  This will render the pool unable to sustain the demands required to maintain cone life and functionality (i.e., nutrient starvation) when $B_2$ becomes negative.  The functionality of the photoreceptor in this model is illustrated by a balance between natural renewal, $a_i$, and shedding, $\mu_i, i=n,m,c$ of the OS.  Physiological experimental results have illustrated a lack of functionality by shorter OS and a larger diameter of the disc at the tip of the OS\cite{Yang}.  Disturbances in the shedding or renewal process can lead to these morphological changes.
 As a result of disturbances on this delicate balance the cones would also eventually die off \cite{Ripps1978}.  The stability of this equilibrium, $E_4$, may give insight into {\it why} the secondary wave of cone death may follow.  Physiologically, we note that, e.g., $D_c=\frac{\mu_c}{a_c}$ can increase with an increase in the shedding rate or a slowing of the renewal rate.  As $D_n, D_m, D_c, D_T$ are all involved in the stability and/or existence of this equilibrium, experiments would need to keep in mind the potential interplay of these quantities in determining exactly what processes may effect slight but crucial differences in these quantities.\\

The eigenvalues of $E_5, E_6, E_7$ are too cumbersome to write explicitly but we can still gain insight by examining where each is physiologically relevant and in relation to the stability of the other equilibria through bifurcations.  We can, however, write the explicit formulation for any $\lambda=0$ bifurcations.  Because our {\it Result} states that the bifurcations describing the progression of RP occur through transcritical bifurcations, these expressions will be useful.\\

\noindent {\bf $E_5= (0,\frac{B_3}{a_md_m}, C_5^*,D_m)$}\\
$E_5$ represents the long term existence of mutated rods, cones, and trophic factors with only the normal rods having disappeared.  This point is physiologically relevant when $C_5^*>0$ and $B_3>0$.  Remaining at this state is desirable for someone with RP, as presumably the patient will still have color and day vision (due to the presence of the cones) and some night vision (provided by the mutated rods).  The eigenvalues are given by $\frac{A_1}{a_m}$ and a complicated cubic polynomial.
Any $\lambda=0$ bifurcations are given by
$$
\frac{\mu_m\gamma A_1B_3C_5^*}{a_m^2}=0.
$$
Equilibrium point $E_5$ corresponds to an earlier stage of the typical rod-cone RP in which only the normal rods have died off and both  cones and mutated rods still remain. We would again expect this equilibrium point to physiologically represent a temporary state before cones and mutated rods eventually die off as parameters in the system change.\\

\noindent {\bf $E_6= (R_{n6}^*,R_{m6}^*,0,D_n)$}\\
This equilibrium represents the death of all cones with the normal and mutated rods as well as the trophic factors remaining.  Besides depending on the renewal and shedding rates, $E_6$ also depends on the rate at which both types of rods remove trophic factors from the trophic pool ($\beta_m, \beta_n$) and the production of the trophic factors itself ($\Gamma$).  This equilibrium can be interpreted as the first stage in the progression of reverse RP in which an individual loses color and day vision.  The eigenvalues are given by
$$
\frac{C_7^* \gamma(d_n A_1+d_m a_n m)}{m\beta_m a_n+\beta_n A_1}
$$
and a complicated cubic polynomial.  Any $\lambda=0$ bifurcations are given by
$$
\frac{(\mu_n+m)\gamma A_1A_3C_7^*}{a_n^2}=0.
$$
In the paths involved in the various progression of RP, $E_6$ loses stability in a transcritical bifurcation with $E_3$ in the sequence 
$E_7\to E_6\to E_3\to E_2$ or with $E_2$ in the sequence $E_7\to E_6\to E_2$.  Similarly, $E_6$ gains stability through a transcritical bifurcation with $E_7$.  Understanding the mechanisms (i.e., the parameters involved) in $E_7\to E_6$ is key as this would give insight into circumventing any vision loss. \\

\noindent {\bf $E_7= (R_{n7}^*,R_{m7}^*,C_7^*,D_n)$}\\
This equilibrium represents the only state in which all photoreceptors continue to survive. It illustrates the first stage in RP where potentially there are no clinical manifestations of RP.  $E_7$, in contrast to $E_6$, besides depending on  parameters governing production of the trophic factors additionally depends on the ability (or lack thereof) of the cones to utilize the RdCVF, quantified through $d_n$ and $d_m$.  The eigenvalues are given by a complicated 4th order polynomial.   The $\lambda=0$ bifurcations are given by
$$
\frac{(\mu_n+m)\gamma A_1B_3C_7^*}{a_n^2}=0.
$$
This is the most desirable state for an individual with RP---both types of rods, the cones, and the trophic factors are all still present in the system.  Through an in depth analysis of this steady state, we can examine how to prevent RP from running its normal course by considering the conditions that will bring about qualitative changes that will render this state unstable and move the system to another steady state previously mentioned.

\section*{Appendix 2: Parameter Estimation}
In obtaining estimates on the parameters for (\ref{full1}), we begin with the healthy eye (non-disease) version
\begin{eqnarray}
\dot{R_n}&=&R_n(a_n T- \mu_n) \nonumber \\
\dot{C}&=&C(a_cT -\mu_c+d_nR_n) \nonumber \\
\dot{T}&=&T(\Gamma-kT-\beta_n R_n-\gamma C).
\label{full0}
\end{eqnarray}
A few key experiments allow us to estimate five of these parameters. We then solve the inverse problem with 9 data points from the aforementioned experiments to estimate the remaining four parameters.  

\noindent\underline{\bf Shedding of photoreceptor outer segments}\\
Numerous experiments, some {\it in vivo} and some in post-mortem patients, have estimated the number of rod OS and cone OS that are shed per day.  This number is typically given in the number of days necessary to regenerate, i.e., all the OS to shed and be replaced by new ones.  Estimates depend on both the location of photoreceptor and the species under consideration.  If we consider only rhesus monkey and humans, we obtain an approximate shedding range (in days) of 
$\mu_n \in \left[\frac{1}{9}, \quad \frac{1}{15}\right],
\mu_c \in \left[\frac{1}{8.5}, \quad \frac{1}{15}\right].
$
See \cite{text,Young,Jonnal,Boulton} for some of these key experiments and results.

\noindent\underline{\bf Rod-derived Cone Viability Factor}\\
The ground breaking experiment of Leveillard et al.~gives insight the approximate value for the help the rods give the cones \cite{Leveillard}.  In their experiment in the {\it rd1} mouse they showed that the presence of RdCVF in a nutrient rich medium resulted in a 40\% rescue effect.  The number of cones per mm$^2$ after one week was 4200 compared with 3200 in their cultured retinal explants.  Assuming an exponential decay to represent the degradation in the absence of RdCVF, we consider the equations
\begin{equation}
\dot{C}=-bC, \qquad \dot{C}=-bC+d_n R_nC
\end{equation}
to represent the cones in the absence and presence, respectively, of RdCVF.  Solving and eliminating $b$ and assuming that we have a level of RdCVF remaining near constant levels (where we let $R^{\mbox{max}}$ represent the number of rods that would be needed to generate this) yields
\begin{equation}
d_n{R_n}^{\mbox{max}} \approx \frac{\ln\left(\frac{42}{32}\right)}{7\cdot 24}.
\end{equation}  
Approximating $R^{\mbox{max}}$ as the average number of rods per mm$^2$ ($87.0\times 10^3$\cite{text,Curcio}) gives a lower bound. 
Knowing the equilibrium value (in terms of the various parameters) for $R_n^*$ in the non-disease model gives another way of calculating $d_n$ as the value is constrained once $a_n, a_c, \mu_n, \mu_c$ are known.

\noindent\underline{\bf Inverse problem}\\
An experiment on rhesus monkeys in which intact inner segments were transplanted into the retina give us estimates on the renewal of rods and cones \cite{Guerin}.  The specific data, which also gave us the renewal estimates, states that the height of the OS discs on the rods (given over mm$^2$ area and as a fraction of the normal height) was .3, .33, and .44 at the respective times of 7, 14, and 30 days after implant.  In the cones, the same fractions of the overall height of the OS were .31, .35, and .48.  At the measurement of 150 days, the height of the implanted photoreceptors was indistinguishable from that of the normal photoreceptors.  Assuming that the observations on this scale hold across the retina and having known estimates of the numbers of rods and cones in the retina, we have 8 data points that are the averaged over the 12 rhesus monkeys from the experiments.   We also know the number of RPE at steady state and this gives us a final value for the trophic pool.  Additionally, the same observations of \cite{Guerin} give observed values of the rod renewal. 

The intervals given represent the range of the four best fit values (using Maximum Likelihood Estimator [MLE]) utilizing the average total number of rods and cones in the human eye as $91\times 10^6$ and $4.5\times 10^6$, respectively \cite{text}.  As there is approximately 1 RPE cell for every 45 photoreceptors, we estimate the total RPE in the retina at $2.1\times 10^6$.  Substitution of these quantities gives us the parameter values of our system; see Table \ref{Table3}.

\begin{figure*}
\includegraphics[width=6 in, height=6 in]{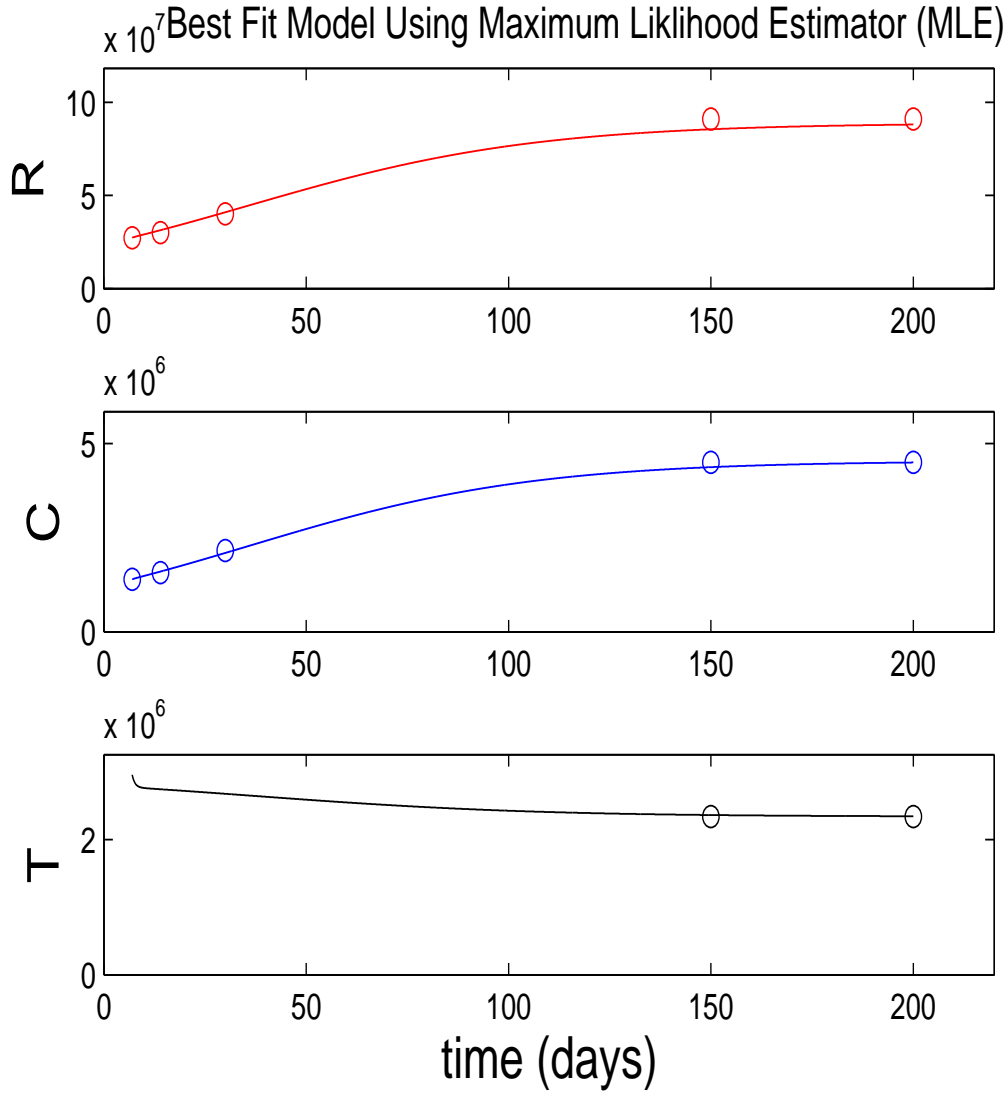}
\caption{ {\bf Best least squares fit (MLE) for inverse problem using data from \cite{Guerin}.}  The quantities in the trophic equation are unknown and all but one of the other parameters have a range of values from the experimental literature. The five circles in each graph are the data points at 7, 14, 30, 150, and 250 days, with only the last two points known in the RPE. The specific values used are $a_n=4.5\times 10^{-8}, a_c=4.7\times 10^{-8}, \Gamma=1.48,  k=5\times 10^{-7},  \gamma=4.88\times 10^{-8},  \mu_n=1/9.5, \mu_c=1/9, \beta_n=1\times 10^{-9},  d_n=1.29\times 10^{-11}. $ The assumed steady state values of $R_n^*\approx 91\times 10^6, C^*\approx 4.5\times 10^6, T^*\approx 2.3 \times 10^6$ for the normal rod OS, cones OS, and RPE, respectively. }\label{InvProb1}
\end{figure*}

The best fit graph is given in Figure \ref{InvProb1}.

For the diseased model (\ref{full1}), we approximate $\beta_m= \beta_n$, $d_m=d_n$, $a_m=a_n$ as first approximations since we have not found data on any of those values.  
To estimate the mutation rate, $m$, we estimated the total number of mutated rods based on the 14 year progression of the disease (patient first noticed symptoms at age 10, and died at age 24, with the postmortem eye analyzed \cite{Szamier1979}.  In doing so, we estimate a mutation rate of $m=3.68\times 10^{-7}/$day.

\section*{Appendix 3: Other Stable Solutions}
\begin{figure*}
\begin{tabular}{p{2.8in}@{}p{2.8in}}
\resizebox{2.8in}{2.5in}
{\includegraphics{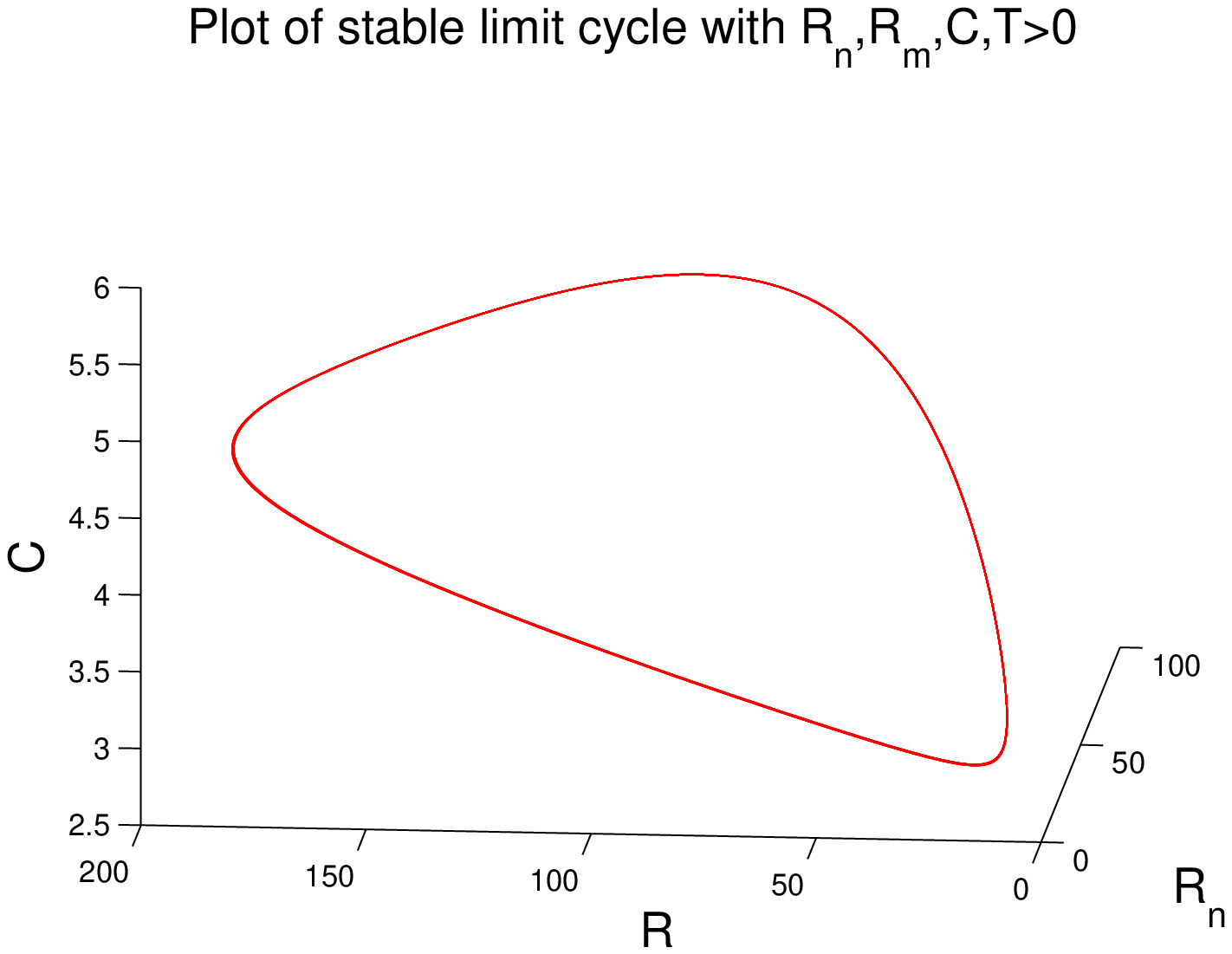}} & 
\resizebox{2.8in}{2.5in}
{\includegraphics{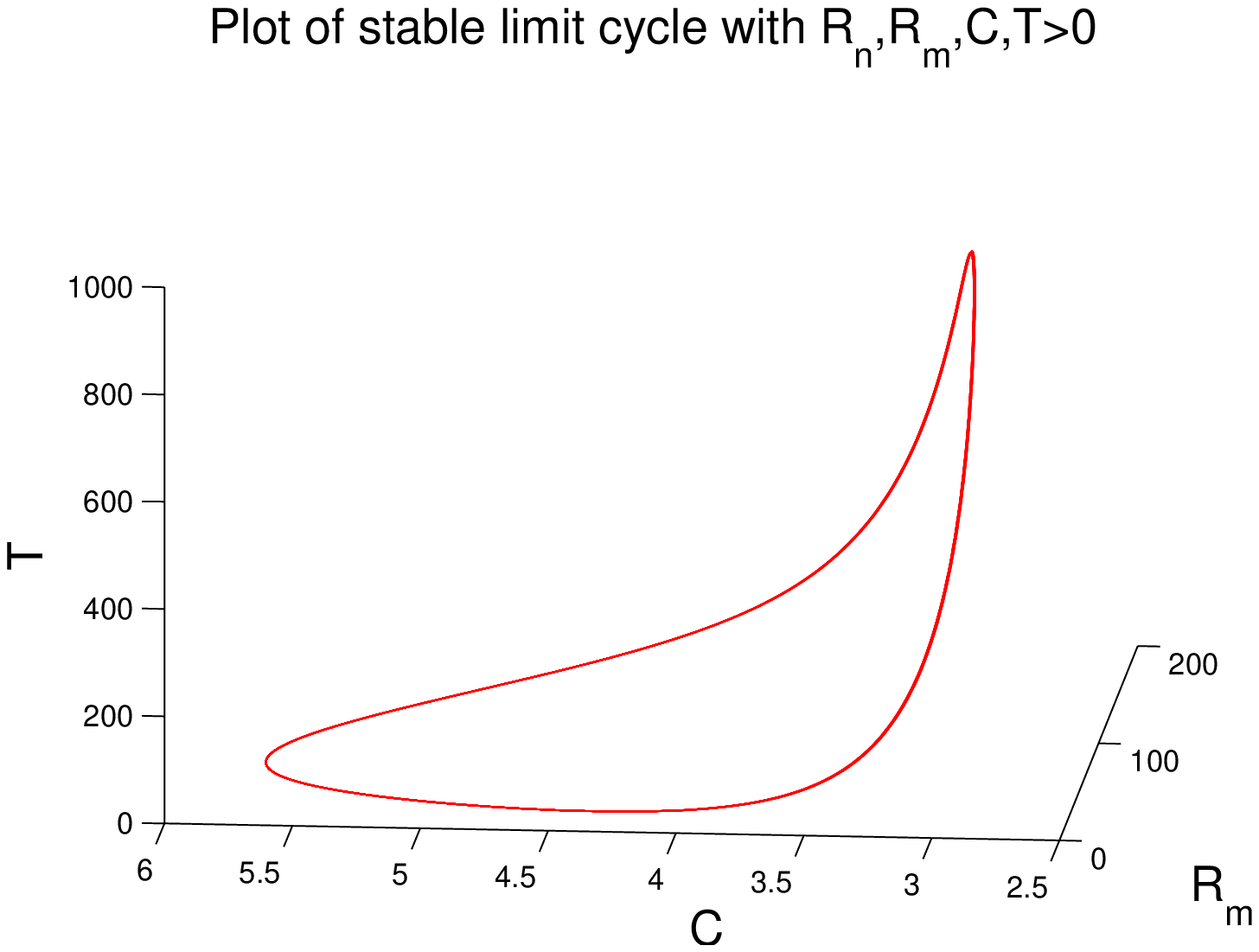}} 
\end{tabular}
\caption{ {\bf Stable limit cycle observed in the full 4-dimensional space.} The parameter values at which this stable limit cycle exists are $\Gamma=1.5$,
$a_n=0.0028$, $a_c=0.00011$, $a_m=0.0027$, $m=0.01$, $\mu_n=0.4$,
$\mu_c=0.15$, $\mu_m=0.4$, $d_n=0.0002$, $d_m=0.00215$, $k=0.0001$, $\beta_n=0.015$, $\beta_m=0.015$, $\gamma=0.037$. Note that the axes are not in millions but are instead the number of OS and RPE (original definition). }
\label{SLC7}
\end{figure*}

\begin{figure*}
\begin{tabular}{p{2.8in}@{}p{2.8in}}
\resizebox{2.8in}{2.5in}
{\includegraphics{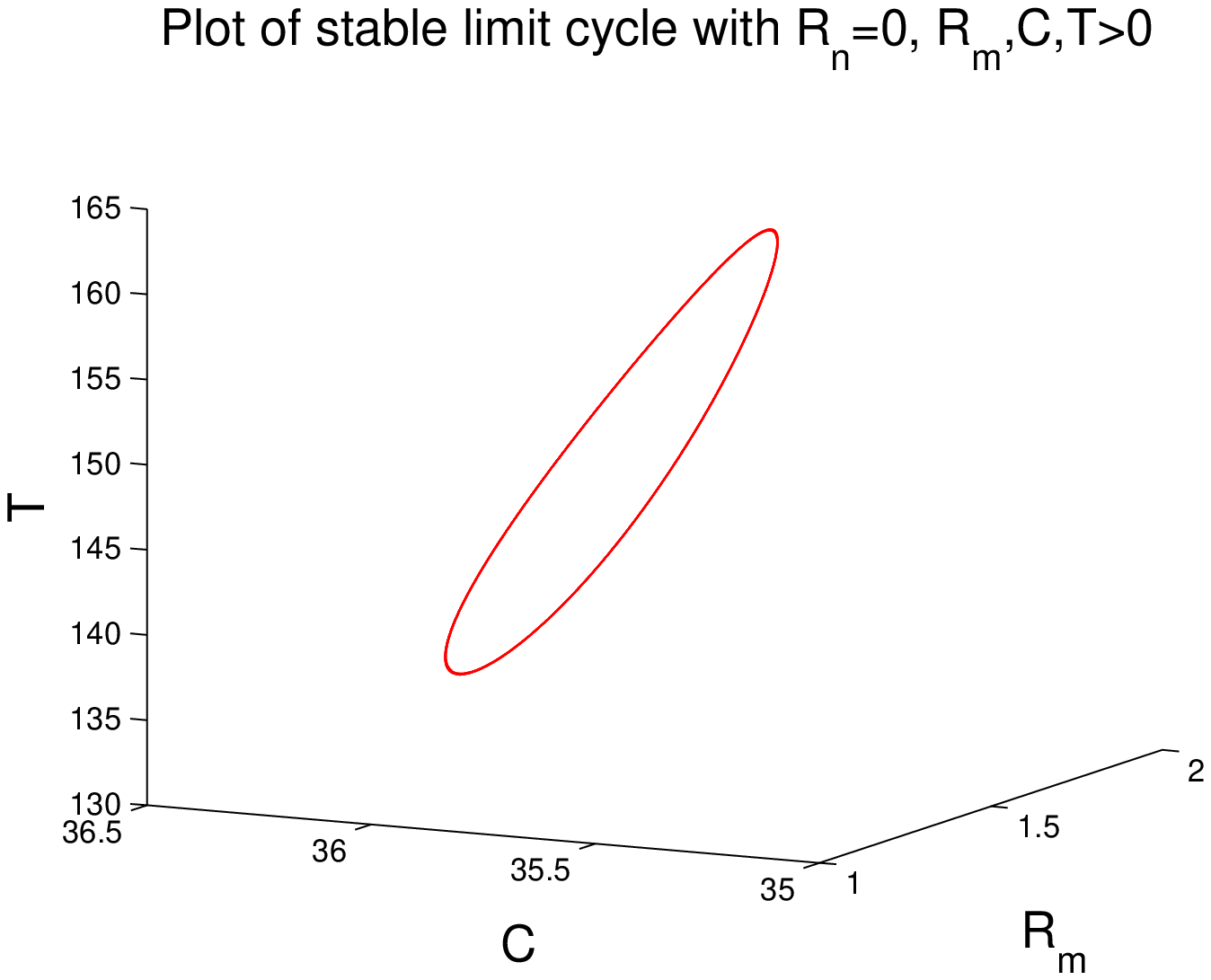}} & 
\resizebox{2.8in}{2.5in}
{\includegraphics{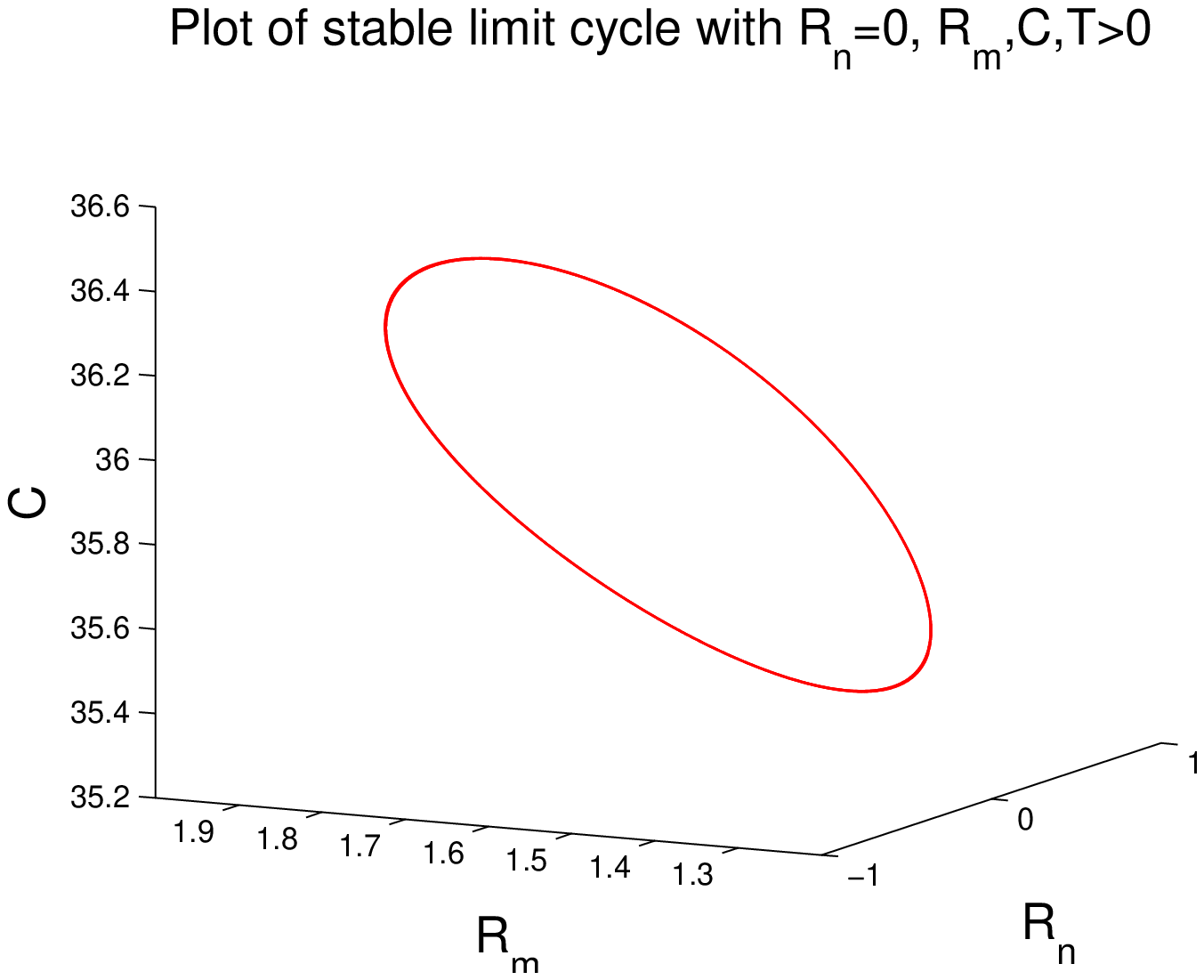}} 
\end{tabular}
\caption{ {\bf Stable limit cycle observed in the 3-dimensional subspace, $R_n=0$.} The parameter values at which this stable limit cycle exists are $\Gamma=1.5$, $a_n=0.0028$, $a_c=0.00011$, $a_m=0.0027$, $m=0.1$, $\mu_n=0.4$,
$\mu_c=0.025$, $\mu_m=0.4$, $d_n=0.0002$, $d_m=0.0055$, $k=0.001$, $\beta_n=0.015$, $\beta_m=0.015$, $\gamma=0.037$. Note that the axes are not in millions but are instead the number of OS and RPE (original definition). }
\label{SLC5}
\end{figure*}

As mentioned previously, we have observed stable periodic solutions for certain (unrealistic) parameter values.  Analytically, we have shown that no non-degenerate Hopf bifurcations ($\lambda=i\omega, \omega>0$) occur for any of the solutions $E_1,\cdots, E_7$ for any choices of parameter values (realistic or unrealistic).  For $\Gamma=1.5$ and the unrealistic values
$a_n=0.0028$, $a_c=0.00011$, $a_m=0.0027$, $m=0.01$, $\mu_n=0.4$,
$\mu_c=0.15$, $\mu_m=0.4$, $d_n=0.0002$, $d_m=0.00215$, $k=0.0001$,
$\beta_n=0.015$, $\beta_m=0.015$, $\gamma=0.037$, a stable limit cycle exists in the full 4-dimensional space; see Figure \ref{SLC7}.  For the same parameter values except for $\mu_c=0.025$, $d_m=0.0055$, $k=0.001$, $m=0.1$, we observe a stable limit cycle in the 3-dimensional subspace $R_n=0$; see Figure \ref{SLC5}.
Following these solutions with MatCont has not yet allowed us to determine the bifurcation that gave rise to either of these limit cycles.

\newpage 


\begin{table}
\caption{Summary of parameters for non-diseased system taken from best 6 MLE results. The equilibrium values for rods, cones, and trophic (RPE) are taken as $91\times 10^6$, $4.5\times 10^6$, $2.3\times 10^6$, respectively. }\label{Table3}
\begin{tabular}{|c|c|c|r|}\hline
Variable & Description & Source \\ \hline
$\Gamma \in [1.42,~1.56]$ & Total inflow rate into the trophic factors & MLE, \cite{Guerin}\\
$k=5\times 10^{-7}$ & Limiting capacity of trophic factors & MLE, \cite{Guerin} \\
$\beta_n= 1.0\times 10^{-9}$ & Constant per cell rate at which $R_n$ withdraws $T$ & MLE, \cite{Guerin}\\
$\gamma\in [4.55,~ 5.44]\times 10^{-8}$ & Constant per cell rate at which $C$ withdraws $T$ & \cite{Guerin}\\
$d_n\in [0.90,~ 1.40]\times 10^{-11}$ & Constant per cell rate at which $R_n$ helps $C$ & \cite{Leveillard,Guerin}\\
$\mu_n\in [\frac{1}{9.5},~ \frac{1}{10}]$ & OS shedding rate of $R_n$ &  \cite{text,Young} \\
$\mu_c\in [\frac{1}{9},~ \frac{1}{9.5}]$ & OS shedding rate of $C$  &  \cite{text,Young,Jonnal} \\
$a_n\in [4.3,~ 4.6]\times 10^{-8}$ & Per cell renewal rod rate & MLE, \cite{Guerin}\\
$a_c\in [4.5,~ 4.8]\times 10^{-8}$ & Per cell cone renewal rate & MLE, \\
\hline
\end{tabular}
\end{table}

\begin{table}
\caption{Summary of parameters for system (\ref{full1}).}\label{Table2}
\begin{tabular}{|c|c|c|r|}\hline
Variable & Description & Units \\ \hline
$\Gamma$ & Total inflow rate into the trophic factors & 1/t\\
$k$ & Limiting capacity of trophic factors & 1/$Tt$ \\
$\beta_n$ & Constant per cell rate at which $R_n$ withdraws $T$ & 1/$R_nt$\\
$\beta_m$ & Constant per cell rate at which $R_m$ withdraws $T$ & 1/$R_mt$\\
$\gamma$ & Constant per cell rate at which $C$ withdraws $T$ & 1/$Ct$\\
$d_n$ & Constant per cell rate at which $R_n$ helps $C$ & 1/$Ct$\\
$d_m$ & Constant per cell rate at which $R_m$ helps $C$ & 1/$Ct$\\
$q_1$ & Per cell conversion factor & $R_n$/$T$\\
$q_2$ & Per cell conversion factor & $R_m$/$T$\\
$q_3$ & Per cell conversion factor & $C/T$\\
$\mu_n$ & Per cell death rate of $R_n$ &  1/$t$  \\
$\mu_m$ & Per cell death rate of $R_n$ &  1/$t$  \\
$\mu_c$ & Per cell death rate of $C$  &  1/$t$ \\
$m$ & Per cell mutation rate of $R_n$ & 1/$t$\\
$a_n$ & Per cell renewal rate $\beta_n q_1$ & 1/$Tt$\\
$a_m$ & Per cell renewal rate $\beta_m q_2$ & 1/$Tt$\\
$a_c$ & Per cell renewal rate $\gamma q_1$ & 1/$Tt$\\
\hline
\end{tabular}
\end{table}

~
\newpage
~

\begin{table}
\caption{Steady states and corresponding stages of RP. Each of the steady state solutions represents a (stable) stage of RP that can be attained for some set of parameter values with the exception of $E_1$, which is never stable.  Many different sub-types of RP can be identified depending on the stable equilibrium that occur as parameters are changed.
}\label{Table1}
\begin{tabular}{|c|c|c|l|}\hline
Steady State & Stage of RP & RP sub-type & Physiological Description\\ \hline
$E_7$ & Early, Mid-late & All types & Normal rods, mutated rods, \\ 
& & & cones, trophic factors all exist\\ \hline
$E_6$ & Mid-late & Cone-Rod & Cones have died off;  both types \\
& & & of rods, trophic factors remain \\ \hline
$E_5$ & Mid & Rod-Cone & No normal rods; mutated rods, \\
& & & cones, trophic factors remain\\ \hline
$E_4$ & Mid-late & Rod-Cone & all rods have died; cones, \\
& & & trophic factors remain\\ \hline 
$E_3$ & Mid-late & Cone-Rod & Cones and normal rods have died;\\
& & & mutated rods and trophic factors remain\\ \hline 
$E_2$ & Late & All types & Complete blindness with no rods or\\
& & & cones; trophic factors remain\\ \hline 
$E_1$ & NA & NA & No rods, cones, or trophic factors\\
\hline
\end{tabular}
\end{table}

\pagebreak 

\end{document}